\documentclass{article}    
\usepackage{latexsym}
\usepackage{amsbsy}
\usepackage{amssymb}
\usepackage[latin1]{inputenc} 

\title{Nonlinear distributional geometry and general relativity}
\author{Roland Steinbauer\footnote{Electronic mail: roland.steinbauer@univie.ac.at;
supported by research grant P12023-MAT of the Austrian Science Foundation} \\
  \\{\small Department of Mathematics, University of Vienna}\\
  {\small Strudlhofg.\ 4, A-1090 Wien, Austria}}
\date{}
\begin{document}
\newcommand{\ep}{\hspace*{\fill}$\Box$}
\newcommand{\eps}{\varepsilon}
\newcommand{\pr}{{\bf Proof. }}
\newcommand{\ms}{\medskip\\}
\newcommand{\cl}{\mbox{\rm cl}}
\newcommand{\g}{\ensuremath{\mathfrak g} }
\newcommand{\gc}{${\cal G}$-complete }
\newcommand{\sa}{\stackrel{\scriptstyle s}{\approx}}
\newcommand{\prol}{\mbox{\rm pr}^{(n)}}
\newcommand{\prolo}{\mbox{\rm pr}^{(1)}}
\newcommand{\deta}{\frac{d}{d \eta}{\Big\vert}_{_{0}}}
\newcommand{\detas}{\frac{d}{d \eta}{\big\vert}_{_{0}}}
\newcommand{\R}{\mathbb R}
\newcommand{\N}{\mathbb N}
\newcommand{\C}{\mathbb C}
\newcommand{\Z}{\mathbb Z}
\newcommand{\K}{\mathbb K}
\newcommand{\sR}{\mathbb R}
\newcommand{\sN}{\mathbb N}
\newcommand{\gK}{{\cal K}} 
\newcommand{\gR}{{\cal R}} 
\newcommand{\gC}{{\cal C}} 
\newcommand{\Dp}{${\cal D}'$ }                                          
\newcommand{\grn}{${\cal G}(\R^n)$ }
\newcommand{\grp}{${\cal G}(\R^p)$ }
\newcommand{\grq}{${\cal G}(\R^q)$ }
\newcommand{\gs}{\ensuremath{{\mathcal G}} }
\newcommand{\gso}{\ensuremath{{\mathcal G}(\Omega)} }
\newcommand{\gsrn}{\ensuremath{{\mathcal G}(\R^n)} }
\newcommand{\gsrp}{\ensuremath{{\mathcal G}(\R^p)} }
\newcommand{\gsrq}{\ensuremath{{\mathcal G}(\R^q)} }
\newcommand{\gst}{\ensuremath{{\mathcal G}_\tau} }
\newcommand{\gsto}{\ensuremath{{\mathcal G}_\tau(\Omega)} }
\newcommand{\gstrn}{\ensuremath{{\mathcal G}_\tau(\R^n)} }
\newcommand{\gstrp}{\ensuremath{{\mathcal G}_\tau(\R^p)} }
\newcommand{\gstrq}{\ensuremath{{\mathcal G}_\tau(\R^q)} }
\newcommand{\es}{\ensuremath{{\mathcal E}} }
\newcommand{\esm}{\ensuremath{{\mathcal E}_M} }
\newcommand{\ns}{\ensuremath{{\mathcal N}} }
\newcommand{\nso}{\ensuremath{{\mathcal N}(\Omega)} }
\newcommand{\est}{\ensuremath{{\mathcal E}_\tau} }
\newcommand{\nst}{\ensuremath{{\mathcal N}_\tau} }
\newcommand{\ks}{\ensuremath{{\mathcal K}} }
\newcommand{\rs}{\ensuremath{{\mathcal R}} }
\newcommand{\cs}{\ensuremath{{\mathcal C}} }
\newcommand{\hsp}{\hspace{-0.3em}}
\newcommand{\comp}{\subset\subset}
\newcommand{\interior}[1]{\mbox{int}(#1)}
\newcommand{\cinfty}{{\cal C}^\infty}
\newcommand{\ck}{{\cal C}^k}
\newcommand{\catensor}{\ensuremath{{\mathcal A}^\otimes} }
\newcommand{\gtn}{{\cal G}_\tau(\R^n) }
\newtheorem{thr}{\hspace*{-3mm} \bf}[section]
\newcommand{\bt}{\begin{thr} {\bf Theorem. }}
\newcommand{\et}{\end{thr}}
\newcommand{\bp}{\begin{thr} {\bf Proposition. }}
\newcommand{\bc}{\begin{thr} {\bf Corollary. }}
\newcommand{\blem}{\begin{thr} {\bf Lemma. }}
\newcommand{\bex}{\begin{thr} {\bf Example. }\rm} 
\newcommand{\bexs}{\begin{thr} {\bf Examples. }\rm}
\newcommand{\bd}{\begin{thr} {\bf Definition. }}
\newcommand{\beast}{\begin{eqnarray*}}
\newcommand{\eeast}{\end{eqnarray*}}
\newcommand{\wsc}[1]{\overline{#1}^{wsc}}
\newcommand{\todo}[1]{$\clubsuit$\ {\tt #1}\ $\clubsuit$}
\newcommand{\rem}[1]{\vadjust{\rlap{\kern\hsize\thinspace\vbox%
                       to0pt{\hbox{${}_\clubsuit${\small\tt #1}}\vss}}}}
\newcommand{\ahat}{\ensuremath{\hat{\mathcal{A}}_0(M)} }
\newcommand{\atil}{\ensuremath{\tilde{\mathcal{A}}_0(M)} }
\newcommand{\aqtil}{\ensuremath{\tilde{\mathcal{A}}_q(M)} } 
\newcommand{\ehat}{\ensuremath{\hat{\mathcal{E}}(M)} } 
\newcommand{\emhat}{\ensuremath{\hat{\mathcal{E}}_m(M)} }
\newcommand{\nhat}{\ensuremath{\hat{\mathcal{N}}(M)} }
\newcommand{\ghat}{\ensuremath{\hat{\mathcal{G}}(M)} } 
\newcommand{\lhat}{\ensuremath{\hat{L}_X} }                    
\newcommand{\SSS}{{\cal S}}
\newcommand{\lla}{\langle}\newcommand{\ra}{\rangle}
\newcommand{\al}{\alpha}
\newcommand{\bet}{\beta} 
\newcommand{\ga}{\gamma}
\newcommand{\Om}{\Omega}\newcommand{\Ga}{\Gamma}\newcommand{\om}{\omega}
\newcommand{\si}{\sigma}\newcommand{\la}{\lambda}
\newcommand{\de}{\delta}
\newcommand{\vphi}{\varphi}\newcommand{\dl}{{\displaystyle \lim_{\eta>0}}\,}
\newcommand{\intl}{\int\limits}\newcommand{\su}{\sum\limits_{i=1}^2}
\newcommand{\D}{{\cal D}}\newcommand{\Vol}{\mbox{Vol\,}}
\newcommand{\Or}{\mbox{Or}}\newcommand{\sign}{\mbox{sign}}
\newcommand{\na}{\nabla}\newcommand{\pa}{\partial}
\newcommand{\ti}{\tilde}\newcommand{\T}{{\cal T}} \newcommand{\G}{{\cal G}}
\newcommand{\DD}{{\cal D}}\newcommand{\X}{{\cal X}}\newcommand{\E}{{\cal E}} 
\newcommand{\CC}{{\cal C}}\newcommand{\vo}{\Vol}
\newcommand{\bat}{\bar t}
\newcommand{\bx}{\bar x}
\newcommand{\by}{\bar y} \newcommand{\bz}{\bar z}\newcommand{\br}{\bar r}
\newcommand{\fr}{\frac{1}}\newcommand{\il}{\int\limits}
\newcommand{\nn}{\nonumber}
\newcommand{\supp}{\mathop{\mathrm{supp}}}

\newcommand{\vp}{\mbox{vp}\frac{1}{x}}\newcommand{\A}{{\cal A}}
\newcommand{\Ll}{L_{\mbox{\small loc}}}\newcommand{\Hl}{H_{\mbox{\small loc}}}
\newcommand{\Lll}{L_{\mbox{\scriptsize loc}}}
\newcommand{\be}{ \begin{equation} }\newcommand{\ee}{\end{equation} }
\newcommand{\beq}{ \begin{equation} }\newcommand{\eeq}{\end{equation} }
\newcommand{\bea}{\begin{eqnarray}}\newcommand{\eea}{\end{eqnarray}}
\newcommand{\beas}{\begin{eqnarray*}}\newcommand{\eeas}{\end{eqnarray*}}
\newcommand{\beqs}{\begin{equation*}}\newcommand{\eeqs}{\end{equation*}}
\newcommand{\lb}{\label}\newcommand{\rf}{\ref}
\newcommand{\GL}{\mbox{GL}}\newcommand{\bfs}{\boldsymbol}
\newcommand{\ben}{\begin{enumerate}}\newcommand{\een}{\end{enumerate}}
\newcommand{\ba}{\begin{array}}\newcommand{\ea}{\end{array}}
\newtheorem{thi}{\hspace*{-1.1mm}}[section]
\newcommand{\bthm}{\begin{thr} {\bf Theorem. }}
\newcommand{\bprop}{\begin{thr} {\bf Proposition. }}
\newcommand{\bcor}{\begin{thr} {\bf Corollary. }}
\newcommand{\bdef}{\begin{thr} {\bf Definition. }}
\newcommand{\brem}{\begin{thr} {\bf Remark. }\rm}
\newcommand{\bth}{\begin{thr}\rm}

\newcommand{\ethi}{\end{thr}}
\newcommand{\id}{\mathop{\mathrm{id}}}
\newcommand{\tr}{\mathrm{tr}}
\newcommand{\spn}{\mathrm{span}}
\newcommand{\lgl}{\langle}
\newcommand{\rgl}{\rangle}
\newcommand{\ca}{{\cal A}}
\newcommand{\cb}{{\cal B}}
\newcommand{\cc}{{\cal C}}
\newcommand{\cd}{{\cal D}}
\newcommand{\cg}{{\cal G}}
\maketitle
\begin{abstract}     
This work reports on the construction of a nonlinear distributional 
geometry (in the sense of Colombeau's special setting) and its applications 
to general relativity with a special focus on the distributional
description of impulsive gravitational waves.
\end{abstract}

\section{Introduction}

Idealizations play an overall role in modeling physical situations; a particularly
useful one is to replace smooth extended densities by ``concentrated sources''
whenever the density is confined to a ``small region'' in space and
its internal structure is negligible (e.g.\ point charges in electrodynamics). 
On trying to describe this idealization mathematically one is led to 
distributions in a natural way. 
In the case of, e.g., electrodynamics distribution theory in fact furnishes a 
consistent framework, i.e., provides the following two features: 
first since Maxwell equations are linear with respect
to sources and fields they make sense within 
distributions and second it is 
guaranteed that (say smooth) charge densities close---in the sense of 
$\D'$-convergence---to, e.g., a point charge produce fields that are close
to the Coulomb field. While the first property allows for a mathematically sound
formulation it is precisely the latter one which renders the
idealization physically sensible.

One would wish for a similar mathematical description  
of concentrated sour\-ces in the theory of general relativity. However, 
its field equations, i.e., Einstein's equations form a (complicated) system 
of {\em nonlinear} PDEs. More precisely, since the spacetime metric and its 
first derivatives enter nonlinearly, the field equations simply cannot be formulated 
for distributional metrics. For a more detailed discussion of the geometrical
aspects and, in particular, weak singularities in general relativity see 
\cite{jamesGWA}.
Despite this conceptual obstacle spacetimes involving an energy-momentum tensor 
supported on a hypersurface of spacetime (so-called thin shells) have long since 
been used in general relativity (see \cite{israel1} for the final formulation 
of this widely applied 
approach). 
The description of gravitational sources supported lower dimensional submanifolds
of spacetime (e.g.\ cosmic strings and point particles), however, is more delicate. 
In fact by a result of Geroch and Traschen
\cite{gt} a mathematically sound and at the same time physically resonable description  
(in the sense of a ``limit consistency'' as discussed in the context of 
Maxwell fields above) explicitly excludes the treatment of sources of the gravitational 
field concentrated on a submanifold of codimension greater than one.

Recently nonlinear generalized function methods have been used to overcome this
conceptual obstacle in the context of such different topics in general
relativity as cosmic strings (e.g.\ \cite{clarke,jamesGWA}), (ultrarelativistic)
black holes (\cite{hb5,ultra,mark}), impulsive gravitational waves (e.g.\
\cite{herbertgeo,geo2}) and signature change (e.g.\ \cite{MaN}). For an overview
see \cite{vickersESI}.

In this work we are going to discuss the recently developed global approach
to nonlinear distributional (in the sense of the special version of
Colombeau's construction) geometry (\cite{diss,ndg,gcfw}) and its applications to 
general relativity. While the following section is devoted to a review of the 
former and, in particular, to generalized pseudo-Riemannian geometry, applications
to the distributional description of impulsive gravitational pp-waves 
will be presented in Section \ref{appl}. 
We shall see that despite the absence
of a canonical embedding of distributions Colombeau's special setting, due to
the fact that the basic building blocks automatically are diffeomorphism invariant 
provides a particularly flexible tool
to model singular metrics in the nonlinear context of general relativity. For 
an introduction into the diffeomorphism invariant {\em full} algebras of generalized
functions of \cite{found} and in particular its global formulation (\cite{vim}),
however, we refer to \cite{gwa_diff_1,gwa_diff_2,gwa_diff_3} in this volume; its
applications to general relativity are discussed in \cite{jamesGWA}.

\section{Generalized pseudo-Riemannian geometry}
In the following we use the notational conventions of \cite{ndg,gcfw}.
The (special) {\em algebra of generalized functions} on the (separable, smooth
Hausdorff) manifold $X$ is defined as the quotient $\gs(X):=\esm(X)/\ns(X)$ of 
the space of moderately growing nets of smooth functions $(u_\eps)_{\eps\in(0,1]}\in
\CC^\infty(X)^{(0,1]}=:{\mathcal E}(X)$ modulo negligible nets, where the respective 
notions of moderateness and negligibility are defined 
(denoting by ${\cal P}(X)$ the space of linear differential operators on $X$) by
\beas
        \esm(X)&:=&\{ (u_\eps)_\eps\in{\mathcal E}(X):\ 
        \forall K\subset\subset X,\ \forall P\in{\cal P}(X)\ \exists N\in\N:
  \\&&
        \sup_{p\in K}|Pu_\eps(p)|=O(\eps^{-N})\}
   \\
        \ns(X)&:=& \{ (u_\eps)_\eps\in\esm(X):\ 
        \forall K\subset\subset X,\ \forall n \in\N_0:\
        \sup_{p\in K}|u_\eps(p)|=O(\eps^{n}))\}\,.
\eeas
\noindent
Elements of $\gs(X)$ are denoted by capital letters, i.e., $U=\cl[(u_\eps)_\eps]=
(u_\eps)_\eps+\ns(X)$. $\gs(\_)$ is a fine sheaf of differential algebras with 
respect to the Lie derivative (w.r.t.\ smooth vector fields) defined by 
$L_\xi U:=\cl[(L_\xi u_\eps)_\eps]$. The spaces of moderate resp.\ negligible 
sequences and hence the algebra itself may be characterized locally, i.e., $U\in\gs(X)$ 
iff $U\circ\psi_\al\in\gs(\psi_\al(V_\al))$ for all charts $(V_\al,\psi_\al)$.
Smooth functions are embedded into $\gs$ simply by the ``constant'' embedding 
$\sigma$, i.e., $\sigma(f):=\cl[(f)_\eps]$, hence $\CC^\infty(X)$ is a faithful 
subalgebra of $\gs(X)$. In the absence of a canonical embedding compatibility 
with respect to the distributional setting is established via the notion of 
{\em association}, defined as follows:
a generalized function $U$ is called associated to $0$, $U\approx 0$, 
if $\int_X u_\eps \mu \to 0$ ($\eps\to 0$) for all compactly supported one-densities 
and one (hence every) representative $(u_\eps)_\eps$ of $U$. The equivalence
relation induced by this notion gives rise to a linear quotient space of $\gs(X)$. 
If $\int_X u_\eps \mu \to w(\mu) $ for some $w\in\D'(X)$ then $w$ is called 
the {\em distributional shadow} (or {\em macroscopic aspect}) of $U$ and we write 
$U\approx w$. Similarly we call a generalized function $U$ 
{\em $k$-associated} to $0$ ($0\leq k\leq\infty$), $U\approx_k 0$, if for all $l\leq k$,
all $\xi_1,\dots,\xi_l\in {\mathfrak X}(X)$ and one (hence any) representative 
$L_{\xi_1}\dots L_{\xi_l}\,u_\eps\,\to\,0$ uniformly on compact sets.    
Also we say that $U$ admits $f$ as {\em $\CC^k$-associated function},
$U\approx_k f$, if for all $l\leq k$, all $\xi_1,\dots,\xi_l \in {\mathfrak X}(X)$ 
and one (hence any) representative $L_{\xi_1}\dots L_{\xi_l}\,(u_\eps-f)\,\to\,0$
uniformly on compact sets. 
Finally, inserting $x\in X$ into $U\in\gs(X)$ yields a well defined element of the 
ring of constants  ${\cal K}$ (corresponding to $\K=\R$ resp.\ $\C$), defined
as the set of moderate nets of numbers ($(r_\eps)_\eps \in \K^{(0,1]}$ with
$|r_\eps| = O(\eps^{-N})$ for some $N$) modulo negligible nets
($|r_\eps| = O(\eps^{n})$ for each $n$).

The $\gs(X)$-module of {\em generalized sections} $\gs(X,E)$ of a vector bundle 
$E\to X$---and in particular the space of {\em generalized tensor fields} 
$\gs^r_s(X)$---is 
defined along the same lines using analogous asymptotic estimates with respect to the 
norm induced by any Riemannian metric on the respective fibers. 
We denote generalized sections by $S=\cl[(s_\eps)_\eps]=(s_\eps)_\eps+\ns(X,E)$.
Alternatively we may describe a section $S\in\gs(X,E)$ by a family  
$(S_\al)_\al=((S^i_\al)_\al)_{i=1}^N$, where $S_\al$ is called the 
{\em local expression} of $S$ with its {\em components}
$S^i_\al:=\Psi^i_\al\circ S\circ\psi_\al^{-1} \in\gs(\psi_\al(V_\al))$ 
($(V_\al,\Psi_\al)_\al$ a vector bundle atlas and $i=1,\dots, N$, with $N$ denoting
the dimension of the fibers) satisfying $S^i_\al(x)\,=\,(\bfs{\psi}_{\al\beta})^i_j
(\psi_\beta\circ\psi^{-1}_\al(x))\,S^j_\beta$ $
(\psi_\beta\circ\psi^{-1}_\al(x))$ for all $x\in \psi_\al(V_\al\cap V_\beta)$, 
where $\bfs{\psi}_{\al\beta}$ denotes the transition functions of the bundle.
Smooth sections of $E\to X$ again may be embedded as constant nets, i.e., 
$\Sigma(s)=\cl[(s)_\eps]$. Since $\CC^\infty(X)$ is a subring of $\gs(X)$,
$\gs(X,E)$ also may be viewed as $\CC^\infty(X)$-module and the two respective module 
structures are compatible with respect to the embeddings. Moreover we have 
the following algebraic characterization of the space of generalized sections
\begin{equation}\label{tensorp}
  \gs(X,E)=\gs(X)\otimes\Ga(X,E)\,,
\end{equation}
where $\Ga(X,E)$ denotes the space of smooth sections and the tensor product is 
taken over the module $\CC^\infty(X)$. 
Compatibility with respect to the classical resp.\ distributional
setting again is accomplished using the concept of ($k$-)association
which carries over from the scalar case by (\ref{tensorp}). 

Generalized tensor fields may be viewed likewise as $\CC^\infty$- resp.\
$\gs$-multilinear mappings, i.e., as $\CC^\infty(X)$- resp.\ $\gs(X)$-modules we have 
\beas
   \gs^r_s(X)&\cong&L_{\CC^\infty(X)}({\mathfrak X}(X)^r,{\mathfrak X}^*(X)^s;\gs(X))\\
   \gs^r_s(X)&\cong&L_{\gs(X)}(\gs^0_1(X)^r,\gs^1_0(X)^s;\gs(X))\,,
\eeas
where ${\mathfrak X}(X)$ resp.\ ${\mathfrak X}^*(X)$ denotes the space of
smooth vector resp.\ covector fields on $X$. In \cite{ndg} many concepts of classical
tensor analysis like e.g.\ Lie derivatives (with respect to both smooth and 
generalized vector fields), Lie brackets, tensor products and contraction have been 
generalized to the new setting and we shall use them freely in the sequel. Moreover
several consistency results with respect to smooth resp.\ distributional
geometry (cf.\ \cite{marsden}) have been etsablished. We now begin to develop the 
basics of a generalized pseudo-Riemannian geometry. 
\bdef\label{gmetric}\begin{itemize}
\item[(i)] 
 A generalized $(0,2)$ tensor field $\hat G\in\G^0_2(X)$ is called a {\em 
generalized Pseudo-Riemannian metric} if it has a representative $(\hat 
g_\eps)_\eps$ satisfying
\begin{itemize}
\item[(a)] $\hat g_\eps$ is a smooth Pseudo-Riemannian metric for all $\eps$, and
\item[(b)] $(\det \hat g_\eps)_\eps$ is strictly nonzero on compact sets, i.e.,
        $\forall K \subset\subset X$ $\exists m\in\N$: 
        $\inf\limits_{p\in K}|\det \hat g_\eps(p)|\geq\eps^m$.
\end{itemize}  
\item[(ii)] 
We call a separable, smooth Hausdorff manifold $M$ furnished with a generalized 
pseudo-Riemann\-ian metric $\hat G$ {\em generalized pseudo-Riemannian manifold}
or {\em generalized spacetime} and denote it by $(M,\hat G)$ or 
merely by $M$. The action of the metric on a pair of generalized vector fields will
be denoted by $\hat G(\Xi,H)$ and $\langle\Xi,H\rangle$, equivalently. 
\end{itemize}
\ethi

Note that condition (b) above is precisely equivalent to invertibility of
$\det\hat G$ in the generalized sense. The {\em inverse metric}
$\hat G^{-1}:= \cl[(\hat g^{-1}_\eps)_\eps]$ is a well defined 
element of ${\gs}^2_0(M)$, depending exclusively on $\hat G\ ($i.e.,\ independent
of the particular representative $(\hat g_\eps)_\eps$ $)$.
Moreover if $\hat G\approx_k g$, where $g$ is a classical $\CC^k$-pseudo-Riemannian 
metric then $\hat G^{-1}\approx_k g^{-1}$.
>From now on we denote the inverse metric (using abstract index notation 
cf.\ \cite{penrose_rindler}, Ch.\ 2) by $\hat G^{ab}$, its components by $\hat G^{ij}$ 
and the components of a representative by $\hat g^{ij}\,_\eps$. Also we shall 
denote the generalized metric $\hat G_{ab}$ by $\hat{ds}^2=\cl[(\hat{ds}_\eps^2)_\eps]$
and use summation convention.

\bexs
\begin{itemize}
\item[(i)] 
A sequence $(\hat g_\eps)_\eps$ of classical (smooth) metrics 
constitutes a representative of a generalized metric if it is moderate and
zero-associated to a classical (then necessarily continuous) metric $g$. 
\item[(ii)]
The metric of a two-dimensional cone was modeled in \cite{clarke} by a 
generalized metric (in the full setting) obtained by using the embedding via
convolution. 
\item[(iii)]
The metric of an impulsive pp-wave will be modeled by a generalized one in Section 
\ref{appl}.
\item[(iv)] Further examples may be found e.g.\ in \cite{hb5,ultra,MaN}.
\end{itemize}
\et

A generalized metric $\hat G$ is non-degenerate in the following sense:
$\Xi\in \gs^1_0(X)$, $\hat G(\Xi,H)=0$ $\forall H\in{\gs}^1_0(M)$ 
$\Rightarrow$ $\Xi=0$. Moreover $\hat G$ induces a $\gs(X)$-linear isomorphism 
${\gs}^1_0(M)\to{\gs}^0_1(M)$ by 
\[ \Xi\mapsto \hat G(\Xi,\,.\,)\,,\]
which---as in the classical context---extends naturally to 
generalized tensor fields of all types. Hence from now on we shall use the 
common conventions on upper and lower indices also
in the context of generalized tensor fields. In particular, identifying
a vector field $\Xi^a\in{\gs}^1_0(M) $ with its metrically equivalent one-form 
$\Xi_a$ we denote its contravariant respectively covariant components
by $\Xi^i$ and $\Xi_i$. A similar convention will apply to representatives.

\bdef\label{gcon}
A {\em generalized connection $\hat D$} on a manifold $X$ is a map
${\gs}^1_0(X)\times{\gs}^1_0(X)\to{\gs}^1_0(X)$ satisfying 
\begin{itemize}
\item[(D1)] $\hat D_\Xi H$ is $\gR$-linear in $H$.
\item[(D2)] $\hat D_\Xi H$ is $\gs(X)$-linear in $\Xi$.
\item[(D3)] $\hat D_\Xi(UH)=U\,\hat D_\Xi H+\Xi(U)H$ for all $U\in\gs(X)$.
\end{itemize}
\ethi

Let $(V_\al,\psi_\al)$ be a chart on $X$ with coordinates $x^i$.
The {\em generalized Christoffel symbols} for this chart are given by the
$(\dim X)^3$ functions $\hat \Ga^k_{ij}\in\gs(V_\al)$ defined by
\[ 
        \hat D_{\pa_i}\pa_j\,=\,\sum\limits_k\hat \Ga^k_{ij}\,\pa_k\,.
\]

We are already in the position to state the ``Fundamental Lemma of 
(\-p\-seu\-do\-)\--Rie\-mannian Geometry'' in our setting.
\bthm
Let $(M,\hat G)$ be a generalized pseudo-Riemannian manifold.\,Then there exists a unique
generalized connection $\hat D$ such that
\begin{itemize}
\item[(D4)] $[\Xi,H]=\hat D_\Xi H-\hat D_H\Xi$ and
\item[(D5)] $\Xi\langle H,Z\rangle=\langle \hat D_\Xi H,Z)\rangle+\langle H,\hat D_\Xi Z\rangle$
\end{itemize}
hold for all $\Xi,\,H,\,Z$ in ${\gs}^1_0(M)$. 
$\hat D$ is called {\em generalized Levi-Civita connection} of $M$
and characterized by the so-called Koszul formula
\beq \label{koszul}
\begin{array}{rcl}
        2\lla \hat D_\Xi H,Z\ra
        &=& \Xi\lla H,Z\ra+H\lla Z,\Xi\ra-Z\lla\Xi,H\ra\\
        && -\lla\Xi,[H,Z]\ra+\lla H,[Z,\Xi]\ra+\lla Z,[\Xi,H]\ra\,.
\end{array}
\eeq
\ethi

As in the classical case from the torsion-free condition (i.e., (D4)) 
we immediately infer the symmetry of the Christoffel symbols of the Levi-Civita 
connection in its lower pair of indices. Moreover, from (D3) and the Koszul 
formula (\ref{koszul}) we derive (analogously to the classical case) the following
\bprop On every chart $(V_\al,\psi_\al)$ we have for the 
generalized Levi-Civita connection $\hat D$ of $(M,\hat G)$ and any vector 
field $\Xi\in{\gs}^1_0(X)$
\[ 
        \hat D_{\pa_i}(\Xi^j\pa_j)
        \,=\,\left(\frac{\pa\,\Xi^k}{\pa x^i}
        +\hat \Ga^k_{ij}\,\Xi^j\right)\,\pa_k\,.
\]
Moreover, the generalized Christoffel symbols are given by
\[
        \hat \Ga^k_{ij}\,=\,\frac{1}{2}\,\hat G^{km}\,
        \left(\frac{\pa \hat G_{jm}}{\pa x^i}+\frac{\pa \hat G_{im}}{\pa x^j}
        -\frac{\pa \hat G_{ij}}{\pa x^m}\right)\,.
\]
\ethi

To be able to state the appropriate consistency results with respect to classical 
resp.\ distributional 
geometry we need to define the action of a classical (smooth) connection $D$ on 
generalized vector fields $\Xi$, $H$ by $D_\Xi H\,:=\,\cl[(D_{\xi_\eps}
\eta_\eps)_\eps]$. Now we have
\bprop Let $(M,\hat G)$ be a generalized pseudo-Riemannian manifold.
\begin{itemize}
\item[(i)] If $\hat G_{ab}=\Sigma(g_{ab})$ where $g_{ab}$ is a classical smooth metric
then we have, in any chart, $\hat \Ga^i_{jk}=\Sigma(\Ga^i_{jk})$ (with $\Ga^i_{jk}$ denoting the 
Christoffel Symbols of $g_{ab}$). Hence for all $H\in{\gs}^1_0(M)$
\[ \hat D_\Xi H\,=\,D_\Xi H\,.\]
\item[(ii)] If $\hat G_{ab}\approx_\infty g_{ab}$, $g_{ab}$ a classical smooth metric,
$\Xi$, $H\in{\gs}^1_0(M)$ and $\Xi\approx_\infty\xi\in{\mathfrak X}(M)$, 
$H\approx\eta\in\D'^1_0(M)\ ($or vice versa, i.e.,  $\Xi\approx\xi\in\D'^1_0(M)$, 
$H\approx_\infty\eta\in{\mathfrak X}(M)\,)$ then
\[ \hat D_\Xi H\,\approx D_\xi\eta\,.\]
\item[(iii)] Let $\hat G_{ab}\approx_k g_{ab}$, $g_{ab}$ a classical $\CC^k$-metric,  
then, in any chart, $\hat\Ga^i_{jk}\,\approx_{k-1}\,\Ga^i_{jk}$.
If in addition $\Xi$, $H\in{\gs}^1_0(M)$, $\Xi\approx_{k-1}\xi\in\Ga^{k-1}(M,TM)$ 
and $H\approx_k\eta\in\Ga^k(M,TM)$ then 
\[ \hat D_\Xi H\,\approx_{k-1}D_\xi\eta\,.\]
\end{itemize}
\ethi

Next we define the generalized Riemann, Ricci, scalar and 
Einstein curvature from an invariant point of view. 
However, all the classical formulae will hold on the level of representatives, i.e., 
all the symmetry properties of the respective classical tensor fields carry over to 
our setting. Moreover, the Bianchi identities hold in the generalized sense.
\bdef \label{cquant}
Let $(M,\hat G)$ be a generalized pseudo-Riemannian manifold with Levi-Civita 
connection $\hat D$.
\begin{itemize}
\item[(i)] The {\em generalized Riemannian curvature tensor}
$\hat R_{abc}\,^d\in{\gs}^1_3(M)$ is defined by
\[      \hat R_{\Xi,H}Z\,:=\,\hat D_{[\Xi,H]} Z-[\hat D_\Xi,\hat D_H]Z\,.
\]
\item[(ii)] The {\em generalized Ricci curvature tensor} 
is defined by $\hat R_{ab}:=\hat R_{cab}\,^c$.
\item[(iii)] The {\em generalized curvature (or Ricci) scalar} 
is defined by $\hat R:=\hat R^a\!_a$. 
\item[(iv)] Finally we define the {\em generalized Einstein tensor} 
by $\hat{\mathfrak G}_{ab}:=\hat R_{ab}-\frac{1}{2}R\hat G_{ab}$.
\end{itemize}
\ethi

The framework developed above opens a gate to a wide range of applications
in general relativity. Definition \ref{gmetric} is capable of modeling a large class
of singular spacetimes while at the same time its (generalized) curvature quantities 
simply may be calculated by the usual coordinate formulae.
Hence we are in a position to mathematically rigorously formulate 
Einsteins equations for generalized metrics. Moreover we have at our disposal 
several theorems (which essentially are rooted in \cite{ndg}, Prop.\ 3) guaranteeing 
consistency with respect to linear distributional geometry resp.\ the smooth setting. 

\bthm Let $(M,\hat G)$ a generalized pseudo-Riemannian manifold with 
$\hat G_{ab}\approx_k g_{ab}$.
Then all the generalized curvature quantities defined above are 
$\CC^{k-2}$-associated to their classical counterparts. 
\ethi

In particular, if a generalized metric $\hat G_{ab}$ is $\CC^2$-associated 
to a  vacuum solution of Einstein equations then we have for the generalized Ricci tensor
\[ 
   \hat R_{ab}\,\approx_0 R_{ab}\,=0\,.
\]
Hence $\hat R_{ab}$ satisfies the vacuum Einstein equations in the sense of 
$0$-association (cf.\ the remarks in \cite{sotonTF}).
\vskip11pt

Generally speaking when dealing with singular spacetime metrics 
in general relativity we may apply the steps of the following
scheme: first we transfer the classically singular metric to 
the generalized setting. This may be done by some ``canonical'' smoothing 
or by some other physically motivated regularization. Of course diffeomorphism
invariance of the procedure employed has to be carefully investigated.
Once the generalized setting has been entered, the relevant curvature quantities 
may be calculated componentwise according to the classical formulae.
All classical concepts carry over to the new framework and one
may treat e.g.\ the Ricci tensor, geodesics, geodesic deviation, etc.\ within this 
nonlinear distributional geometry. Finally one may use the concept of ($k$)-association
to return to the distributional or $\CC^k$-level for the purpose of interpretation.

This program has been carried out for a conical metric (representing a cosmic string) 
by Clarke, Vickers and Wilson (\cite{clarke,jamesGWA}, however, in the full 
setting of Colombeau's construction) rigorously assigning to it a distributional 
curvature and (via the field equations) the heuristically expected energy-momentum 
tensor. In Section \ref{appl} we are going to review the distributional 
description of impulsive pp-wave spacetimes of \cite{geo,geo2,penrose}. Further
applications following the proceddure described above may be found e.g.\ in
\cite{hb5,mark}.

\section{Impulsive gravitational waves}\label{appl}
Plane fronted gravitational waves with parallel rays (pp-waves) are
spacetimes charac\-teri\-zed by the existence of a covariantly constant null vector field,
which can be used to write the metric tensor in the form 
\[
  ds^2\,=\,h(u,x,y)du^2-du\,dv+dx^2+dy^2\,
\]
where $u,\,v$ is a pair of null coordinates ($u=t-z$,\, $v=t+z$)
and $x,\,y$ are transverse (Cartesian) coordinates.
We are especially interested in impulsive pp-waves as introduced by R.
Penrose (see e.g.\ \cite{penrose2}) where the profile function $h$ is proportional 
to a $\de$-distribution, i.e., takes the form $h(u,x,y)\,=\,f(x,y)\,\de(u)$,
where $f$ is a smooth function of the transverse coordinates.
This metric is flat everywhere except on the null hypersurface $u=0$, where it has 
a $\de$-shaped ``shock'' and---due to the appearance of a distribution in one 
component---clearly lies beyond the scope of linear distributional geometry.        

Physically this form of the metric arises as the impulsive limit of a sequence
of sandwich waves, i.e., $h_\eps$ taking the form $h_\eps(u,x,y)=\de_\eps(u)f(x,y)$
with $\de_\eps\to\de$ weakly.
This is our motivation to model the impulsive pp-wave metric by a generalized
metric of the form
\beq\label{gm}
   \hat{ds}^2=f(x,y)D(u)du^2-dudv+dx^2+dy^2\,,
\eeq
where $D$ denotes a generalized delta function which allows for a strict delta
net $(\rho_\eps)_\eps$ as a representative, i.e.,
\beas
\mbox{(a)}
&\quad& \mbox{\rm supp}(\rho_\eps) \to \{0\} \quad (\eps\to 0)\,,
\nn\\
\mbox{(b)}
&\quad& \int \rho_\eps (x) \,dx \to 1 \quad (\eps\to 0)\,\,\mbox{ and}\nn\\
\mbox{(c)}
&\quad& \exists \eta>0\,\,\exists C\ge 0: \int |\rho_\eps (x)| \,dx
\le C \,\, \forall  \eps\in (0,\eta]\,. \nn
\eeas  

In \cite{geo2} it has been shown that the geodesic as well as the geodesic deviation
equation for the metric (\ref{gm}) may be solved 
uniquely within our present setting. Moreover
these unique generalized solutions possess physically reasonable distributional
shadows which shows that we have achieved a physically sensible distributional
description of impulsive pp-waves. Diffeomorphism invariance of these results is
assured by diffeomorphism invariance of the class of strict delta nets.

Here, however, we shall be interested in modeling 
the heuristically motivated singular transformation of the distributional pp-wave 
metric first given by R. Penrose (\cite{penrose2}) within our framework (cf.\
\cite{penrose}). For later use we introduce the notation  $X^i(x^j_0,u)$,
$V(v_0,x^j_0,u)$ ($i,j=1,2$) for the unique generalized geodesics of (\ref{gm})
with vanishing initial speeds. Here the fourth coordinate $u$---due to the special 
geometry---may be used as an affine parameter along the geodesics and the real constants
$x^i_0,v_0$ denote the initial positions, i.e., $X^i(x^j_0,-1)=x^i_0$ and 
analogously for $V$.

In the literature impulsive pp-waves have frequently been described in different
coordinates where the metric tensor is actually continuous, i.e., (in the special
case of a plane wave ($f(x,y)=x^2-y^2$ and $u_+$ denoting the kink function),
\begin{equation}\label{cm}
        ds^2\,=\,(1+u_+)^2dX^2+(1-u_+)^2dY^2-dudV\,.
\end{equation}

Clearly a transformation relating these two metrics cannot
even be continuous, hence in addition to involving ill-defined
products of distributions it changes the topological
structure of the manifold. In the special case envisaged above this discontinuous
change of variables was given in \cite{penrose2} (denoting by $H$ the Heaviside 
function)
\bea   x&=&(1+u_+)X,\qquad y\,=\,+(1-u_+)Y\nn\\
        v&=&V+\frac{1}{2}X^2(u_++H(u))+\frac{1}{2}Y^2(u_+-H(u))\,.\label{pt}
\eea   
However, the two mathematically distinct spacetimes are equivalent from a
physical point of view, i.e., the geodesics and the particle motion
agree on a heuristic level (see \cite{bis-proc}). We are now going to model
this transformation by a generalized coordinate transformation, that is

\bd\label{gct}
Let $\Om$ be an open subset of $\R^n$. We call $T\in\G(\Om,\R^n)$ a
{\em generalized diffeomorphism} 
if there exists $\eta> 0$ such that
\begin{itemize}
\item[(i)]  There exists a representative $(t_\eps)_\eps$ such that
$t_\eps:\Om\to t_\eps(\Om)$ is a diffeomorphism for all $\eps\leq\eta$  and
there exists $\tilde\Om\subseteq\R^n$ open, $\tilde\Om\subseteq\bigcap_{\eps\leq\eta}
t_\eps(\Om)$.
\item[(ii)] $(t^{-1}_\eps)_\eps\in\E_M(\tilde\Om,\R^n)$ and there exists $\Om_1\subseteq\R^n$
open such that $\Om_1$ $\subseteq$ $\bigcap_{\eps\leq\eta}t^{-1}_\eps(\tilde\Om)$.     
\item[(iii)] Writing $T^{-1}:=\cl[(t^{-1}_\eps|_{\tilde\Om})_\eps]$, $T\circ T^{-1}$ as
well as $T^{-1}\circ T|_{\Om_1}$ are well-defined elements of $\G(\tilde\Om,\R^n)$ resp.\ $\G(\Om_1,\R^n)$.
\end{itemize}
\et
It is then clear that $T\circ T^{-1}=\mathrm{id}_{\tilde\Om}$ resp.\ $T^{-1}\circ
T|_{\Om_1}=\mathrm{id}_{\Om_1}$.  
Let us now consider the transformation
$T=\cl[(t_\eps)_\eps]: (u,v,x^i)\mapsto (u,V,X^i)$ 
depending on the regularization parameter $\eps$ according to
\begin{equation}\label{trafo}
  \begin{array}{rcl}
    t_\eps:\quad    x^i&=&x^i_\eps(X^j,u)\\[0.2em]
                        v&=&v_\eps(V,X^j,u)\,,
  \end{array}
\end{equation}             
where $x^i_\eps(x^j_0,u)$ and $v_\eps(v_0,x^j_0,u)$ are representatives of the 
generalized geodesics $X^i(x^j_0,u)$ resp.\ $V(v_0,x^j_0,u)$ and are given by
\begin{equation} \label{strsf}
\begin{array}{rcl}
 x^i_\eps(x^k_0,u)&=&x_0^i+\frac{1}{2}\intl_{-\eps}^u\intl_{-\eps}^s\
        \pa_if(x^j_\eps(x_0^k,r))\rho_\eps(r)\,dr\,ds\\[0.2em]
         v_\eps(v_0,x^k_0,u)&=&v_0+\intl_{-\eps}^u f(x^j_
        \eps(x_0^k,s))
          \rho_\eps(s)\,ds \\[0.2em]
          && + \intl_{-\eps}^u\intl_{-\eps}^s\pa_i f(
          x^j_\eps(x^k_0,r))\dot x^i_\eps(x_0^k,r))\rho_\eps(r)\,dr\,ds\,.
\end{array}
\end{equation}   

Now one may prove the following
\bt
The generalized function $T$ defined above is a generalized 
coordinate transformation on a suitable open subset $\Om$ of $\R^4$ containing the
shock hyperplane at $u=0$. The new coordinates are constant along the 
geodesics given by (\ref{strsf}). Moreover the macroscopic apsect of $T$
coincides with the discontinous transformation used in the literature
(hence in the special case of a plane wave is given by (\ref{pt})).
\et    
 
Applying this generalized transformation to the metric (\ref{gm})
we find that in the new coordinates it is given by the class of
\begin{equation}\label{trsfm}
     \hat{ds}_\eps^2\,=\,
        -dudV+(2\sum\limits_{i=1}^2(\dot x^i_\eps\pa_jx^i_\eps)-\pa_jv_\eps)
        dudX^j
        +\sum\limits_{i=1}^2(\pa_jx^i_\eps dX^j)^2\,,
\end{equation}
where
$\dot{\,}$ and $\pa_i$ denote derivatives with respect to $u$ and $X^i$,
respectively. Moreover we find the following distributional shadow
of the metric (\ref{trsfm})
\begin{equation}\label{dslim}
\begin{array}{l}
\cl[(\hat{ds}_\eps^2)_\eps]
    \approx-du dV +(1 + \frac{1}{2} \pa_{11}f(X^j)u_+)^2 dX^2\\[0.3em] +
            (1 + \frac{1}{2} \pa_{22}f(X^j)u_+)^2 dY^2
     + \frac{1}{2} \pa_{12}f(X^j)\triangle f(X^j) u_+^2 dX dY\\[0.3em]    
     + 2 u_+ \pa_{12}f(X^j) dX dY
    +\frac{1}{4} (\pa_{12}f(X^j))^2 u_+^2 (dX^2\! +\! dY^2)\,
\end{array}
\end{equation}    
which is precisely the continuous (or so-called Rosen-) form of the metric 
of an impulsive pp-wave (cf.\ \cite{ab3,pv1}).
\vskip11pt

Summing up we have shown the following: after modeling the (distributional
form of the) impulsive pp-wave metric in a diffeomorphism invariant 
way by the generalized metric (\ref{gm}) we have subjected the latter to
the generalized change of coordinates $T$. In either coordinates
the distributional shadow is computed giving the distributional resp.
the continuous form of the pp-wave metric. (Note that although the action
of a smooth diffeomorphism is compatible with the notion of association,
generalized coordinate transformations clearly are not.)
Physically speaking the two forms of the impulsive  metric  arise  
as the (distributional) limits of a sandwich wave in different coordinate systems.   
Hence impulsive pp-waves indeed are sensibly modeled by the generalized spacetime 
metric (\ref{gm}): in different coordinate systems related by generalized 
coordinate transformations, different distributional pictures arise.

\end{document}